\documentclass{aa}
\usepackage{lineno}

\usepackage{graphicx}
\usepackage{txfonts}
\begin{document} 

   \title{The puzzling new class of variable stars in NGC 3766 : old friend pulsators?}


   \author{S. J. A. J. Salmon
          \inst{1}
          \and
          J. Montalb\'an
          \inst{1}
          \and
           D. R. Reese
          \inst{1,3}
          \and 
          M.-A. Dupret
          \inst{1}
          \and 
          P. Eggenberger
          \inst{2}
          }

   \institute{D\'{e}partement d'Astrophysique, G\'{e}ophysique et Oc\'{e}anographie, Universit\'{e} de Li\`{e}ge, All\'{e}e du 6 Ao\^{u}t 17, 4000 Li\`{e}ge, Belgium
\email{salmon@astro.ulg.ac.be}
          \and
              Observatoire de Gen\`{e}ve, Universit\'{e} de Gen\`{e}ve, Chemin des Maillettes 51, 1290 Sauverny, Switzerland
           \and
               School of Physics and Astronomy, University of Birmingham, Edgbaston, Birmingham, B15 2TT, UK\\
             }

   \date{Received September 15, 1996; accepted March 16, 1997}


       \abstract
   {The recent variability survey of the \object{NGC 3766} cluster revealed a considerable number of periodic variable stars in a region of the Hertzsprung-Russell diagram where no pulsations were expected. This region lies between the instability strips of the $\delta$ Scuti and Slowly Pulsating B (SPB) stars. Moreover the periods of the new phenomenon, P$\sim$0.1--0.7 d, do not allow us to associate it \emph{a priori} to either of these two types of pulsations.}
   {Stars in the \object{NGC 3766} cluster are known to be fast rotators with rotational velocities typically larger than half of their critical velocity. Rotation can affect both the geometrical properties and period domain of pulsations. It also alters the apparent luminosity of a star through gravity darkening, effect seldom taken into account in theoretical studies of the rotation-pulsation interaction. We explore whether both of these effects are able to deliver a consistent interpretation for the observed properties of the ``new variables'' in NGC 3766: that is, explaining their presence outside the instability strips of known pulsators and the domain of their variability periods.}
   {We carry out an instability analysis of SPB models within the framework of the Traditional Approximation of Rotation. We then study the visibility of excited modes according to the angle of view and rotation. We also check how gravity darkening affects the effective temperature and luminosity of stellar models for different angles of view and rotational velocities. We adopt the simple approach of von Zeipel to express gravity darkening.}
   {At the red (cold) border of the instability strip, prograde sectoral modes, which are equatorially trapped waves, are preferentially excited and their visibilities are maximum when seen equator-on. From a linear computation, the amplitudes of the prograde sectoral modes are at best $\sim$40\% of their non-rotating counterparts. This ratio qualitatively reproduces the properties of the variability amplitudes observed between the SPBs and new variables in \object{NGC 3766}. Furthermore low-mass SPB models seen equator-on can appear in the gap between non-rotating SPB and $\delta$ Scuti stars due to gravity darkening. In that case, the periods of most visible modes are shifted to the 0.2--0.5 d range due to the effects of the Coriolis force. We hence suggest that the new variable stars observed in \object{NGC 3766} are actually fast rotating SPB pulsators.}
   {}

   \maketitle
%

\section{Introduction}

   Slowly Pulsating B (SPB) stars are mid- to late B-type objects displaying high-order gravity (g) modes with periods between 0.5 and 5~d \citep[e.g.][]{decatrev}, excited by the $\kappa$ mechanism due to the iron-group opacity bump at $\log T$$\sim$5.3. $\delta$ Scuti stars are late A- and early F-type stars. They show low-order pressure (p) and g modes with periods between 0.3 and 6~hr, activated by the $\kappa$ mechanism, in this case due to the second He ionisation zone at $\log T$$\sim$4.6. The opacity bumps responsible for the $\kappa$ mechanism must be optimally located inside stars in order to work efficiently \citep{pami99}. In early A-type stars the two previously mentioned opacity peaks are located, respectively, too deep and too close to the surface. This is at the origin of a gap between the SPB and $\delta$~Scuti instability strips in the Hertzsprung-Russell (HR) diagram. 

   Very recently \citet[][hereafter Mo13]{mowlavi} analysed, based on a multi-band photometric 7 year survey, the variability of stars in the \object{NGC 3766} cluster. They detected 13 SPB and 14 $\delta$ Scuti candidates. These pulsators correspond to stars on the Main Sequence (MS), in accordance with the age of \object{NGC 3766}, estimated between 10 and 31 Myr \citep[][and references therein]{aidelman}. However, Mo13 surprisingly discovered 36 additional mono- and multi-periodic variable stars spanning over 3 magnitudes, all located in the HR diagram between the detected SPB and $\delta$~Scuti stars. These uncategorised stars present a periodicity between 0.1 and 0.7~d, with a peak near $\sim$0.3~d, clearly outside the respective pulsation domains of SPBs and $\delta$~Scuti. In the visible bandpass, the amplitudes of their variability are, for all but one, below 5~mmag: that is, two to three times smaller than the amplitudes of the SPBs observed in the same cluster. This led the authors to conclude that these objects might constitute a peculiar new class of periodic variable stars. 

    The presence of pulsators between the SPB and $\delta$~Scuti stars has been often debated in the literature. The first to suggest their existence was \citet{struve} who reported unexpected short-period variability in the late-B Maia star, member of the Pleiades, and in one A star. Although the same author had to dismiss the presence of variability in Maia \citep{struve2}, the name of this star is still used to refer to these putative variables.

Meanwhile, several claims of Maia candidate detection were finally dismissed \citep[see for instance][]{decatmaia}, although a population of field stars observed by the CoRoT mission appears now as good Maia candidates. Indeed, \citet{degrootecorot,degroote09coast} detected in the CoRoT field a set of late B-type stars in the gap between SPB and $\delta$ Scuti stars, that show low-amplitude variability with periods rather characteristic of $\beta$~Cephei stars, leading the authors to give the name of ``low-amplitude $\beta$~Cephei stars'' to these objects. \citet{saesenpersei,saesen2} also found in the NGC~884 cluster this kind of late-B type, low-amplitude and low-period pulsators, albeit only one of them fell clearly outside the classic SPB instability strip.

  
   A spectroscopic study of 38~B~stars belonging to \object{NGC 3766}, including 12~Be~stars, showed that more than half of them rotate at more than 50\% of their critical velocity \citep[$v\sin i/v_{\textrm{crit}}$>0.5,][]{mcswain}. Mo13 pointed out that the presence of this new type of variability might be linked to fast rotation in stars of the cluster, thereby suggesting a link with the low-period modes predicted by \citet[][hereafter T05]{townsend} for rotating SPB stars.

   T05 analysed the effects of rotation on pulsation properties of late B-type stars, showing that the periods of g modes can be significantly shifted downwards. Yet, their instability domain remains almost unchanged in effective temperature ($T_{\textrm{e}}$) and luminosity ($L$). Unfortunately, the form under which his results are presented does not lend itself to a direct comparison with observations. \citet{savonije} and \citet{townsendretro} focused their study on Rossby waves, which have no counterpart in the non-rotating case, and appear to be excited in SPB models when rotation is included. These modes extend the red border of the SPB instability strip while their periods are in the range of $\beta$ Cephei stars. They are hence possible candidates to explain the variability detected by Mo13. Nevertheless they might be difficult to observe due to their geometrical properties, as will be discussed in this work. 

   Rotation also affects the star by distorting its surface, leading to a modification of the local emitted radiative flux, as theorised by \citet{vonzeipel}. It is well known that the observed flux of a rotating star will then depend on the angle of view \citep[e.g.][]{maederpeyt}. We wonder how this effect could displace the apparent position of actual SPB stars towards cooler and fainter regions. 

   Therefore we suggest in the present work that the new variable stars detected by Mo13 are in fact fast rotating SPBs. To assess our explanation, we study the effects of rotation on models at the red border of the SPB instability strip:
\begin{itemize}
 \item at first by verifying whether rotation can both shift periods to an unusual range and reproduce the observed properties of the variability amplitudes. 
 \item next, by including gravity darkening in the models and exploring its consequences on $T_{\textrm{e}}$ and $L$ for different rotational velocities and inclinations.
\end{itemize} 

 The paper is divided as follows: in the first two sections we analyse respectively the instability and visibility of modes using the Traditional Approximation of Rotation. In the third section we present the effects of gravity darkening on SPB models for different rotational velocities and angles of view. We end with a discussion and conclusion.

\section{Pulsational instability in the frame of the Traditional Approximation of Rotation}
\label{section1}
The Traditional Approximation of Rotation (TAR) was introduced by \citet{eckart} in geophysics to study the effects of rotation on low-frequency gravity modes. It was independently developed  by \citet{berthomieu} for similar modes in rotating stars. The TAR assumes that the rotational frequency, $\Omega$, is moderate in comparison to the critical rotation rate of the star, or equivalently that the sphericity of the star is conserved. In this work, we adopt the following definition of the critical rotation rate, $\Omega_{\textrm{crit}}=(GM/R^3_{\textrm{e}})^{1/2}$, where $G$ is the gravitational constant, $M$ the mass of the star and $R_{\textrm{e}}$ the radius at the equator.

\begin{table*}
\caption{Instability domains for the SPB models aged $\sim$20 Myr. Each row presents the inertial period domains (in days) and the number of unstable modes (in brackets) found in a model at a given mass and $\Omega$. The equatorial velocity, v$_{\textrm{eq}}$, is also indicated.}             
\label{table1}      
\centering          
\begin{tabular}{l l c c c  c c c c  }     
\hline\hline       
Mass & $\Omega/\Omega_{\textrm{crit}}$ & v$_{\textrm{eq}}$&  \multicolumn{2}{c}{$m=-\ell$} & $-\ell<m<0$ & \multicolumn{2}{c}{$m=0$} & $m=\ell$   \\ 
(M$_{\odot}$) & & (km/s)&$\ell=1$ & $\ell=2$ & $\ell=2$ & $\ell=1$ & $\ell=2$ & $\ell=1$ \\
\hline  
2.9 & 0.20 & 109 & 0.43 - 0.53 (7)  & -- & -- & -- & -- & -- \\ 
 & 0.40 & 218 & 0.29 - 0.33 (7)  & -- & -- & -- & -- & -- \\ 
 & 0.60 & 326 & 0.22 - 0.24 (7)  & -- & -- & -- & -- & -- \\ 
3.2 & 0.20 & 111 & 0.43 - 0.58 (10) & 0.23 - 0.28 (8)   & 0.29 - 0.35 (6) & 0.57 - 0.84 (10) & 0.41 - 0.45 (3) & 0.87 - 1.23 (6)\\ 
 & 0.40 & 223 & 0.30 - 0.35 (9) & 0.15 - 0.18 (9)  & 0.21 - 0.22 (2) & 0.46 - 0.58 (7) & -- & -- \\ 
 & 0.60 & 334 & 0.22 - 0.25 (9) & 0.12 - 0.13 (9)  & -- & 0.40 - 0.45 (4) & -- & -- \\ 
\hline                  
\end{tabular}
\end{table*}

The TAR next proceeds by neglecting the latitudinal components of the Coriolis force (terms in $\sin \theta$, with $\theta$ the colatitude) in the momentum equation. This first approximation is valid for modes whose horizontal displacement is much larger than the radial one, a condition respected by high-order gravity modes in SPB stars. To allow the separability of the oscillation equations under non-adiabaticity, a further approximation is made in the energy equation. This latter is justified by the fact that horizontal radiative flux terms are small for low degree g modes. We refer to \citet{bouabid}, and references therein, for a complete description of these approximations.

\begin{figure}
\centering
\resizebox{ \linewidth}{!}
{\includegraphics{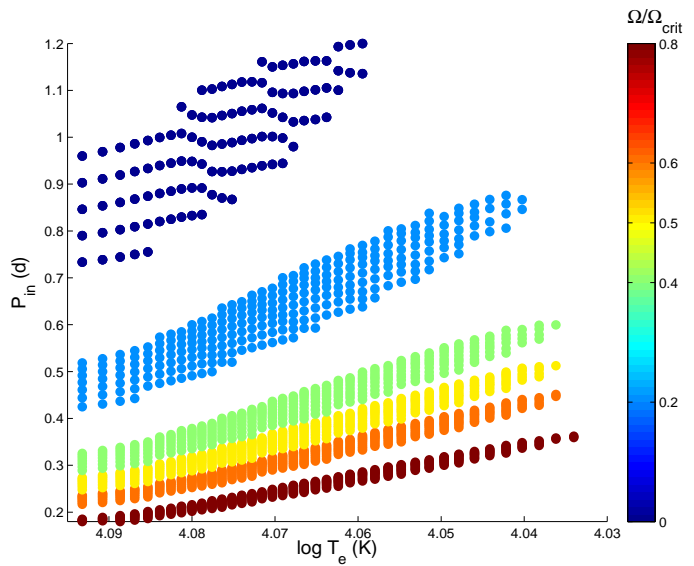}}
\caption{ Inertial period of unstable modes in the 2.9 M$_{\odot}$ model during its evolution on the MS, represented through the evolution of its effective temperature. The colours distinguish the different rotation rates we consider. Their values are reported in the legend to the right. Only $\ell_1$ PS modes are destabilised, regardless of the rotation rate.}
\label{figure1}
\end{figure}
Several authors implemented the TAR to look at the consequences of rotation on SPB pulsations. T05 detailed the evolution of the periods of classic g modes with rotation. He showed that prograde sectoral modes slightly extend the red border of the SPB instability strip to lower temperatures as rotation increases. These latter behave as Kelvin waves and correspond to $\ell=-m \ $ g modes, where $\ell$ is the angular degree and $m$ the azimuthal order. On the contrary, the red borders of retrograde and axisymmetric gravito-inertial modes ($-\ell<m\leq\ell$, Poincar\'e waves) are shifted towards hotter regions. Yet, the presentation of the main results in terms of the period in the corotating frame does not allow us to directly use T05 for investigating the scenario we suggest, and compare it to observations.  

\citet{savonije} and \citet{townsendretro} focused their study on Rossby and Yanai modes, which have no counterpart in the spectra of non-rotating stars (see details in \citealt{leehough}) and showed how they depend on the stellar rotation rate and evolutionary stage. They can present periods smaller than 1~d, especially if ones considers $m>1$. The latter author also found that these modes are excited at lower temperatures than classic SPB modes, still considering the highest $m$. However, following the approach suggested in \citet{savonije2}, we will show in the Discussion section that these modes have low visibilities and hence may be difficult to observe. We note that the former author also analysed, as in T05, the behaviour of Kelvin waves. But he only considered stellar models well inside the SPB instability strip, hence not relevant for the part of the HR diagram we wish to explore.

In a comparison between the TAR and a complete rotation treatment of spherical models, \citet{ballottar} showed that the two approaches agree, even well within the inertial regime. This regime corresponds to $\eta$>1, with $\eta=2\Omega/\omega_{\textrm{co}}$, the spin parameter and $\omega_{\textrm{co}}$ the angular pulsation frequency. With help of non-spherical models, they found that the TAR underestimates the periods of modes, as expected, when the centrifugal distortion starts to play an important role at rapid rotation rates. This underestimation reaches up to 4\% for $\eta \simeq 5$. In this work, we are looking at effects that could shift periods from $\sim$1 to $\sim$0.3~d, i.e. lowering it by 70\% of its initial value. The results provided by TAR computations should thus remain of sufficient precision for our purposes.

\citet{lee08} analysed the stability of g modes using computations carried out with the TAR and a 2D treatment including the rotational distortion effects, respectively. While the retrograde modes ($m$>0), found to be excited within the TAR framework became stable when considering the 2D approach, the instability domain of prograde modes ($m$<0) was to the contrary increased.

Bearing these limits in mind, we determine for different $\Omega$ the excitation and pulsation periods of stars at the red border of the classic SPB instability strip. In our approach, we pay special attention to the properties of pulsations in this region of the HR diagram that would allow for a good interpretation of the Mo13 results.

We start by computing stellar models of respectively 2.8, 2.9, and 3.2~M$_{\odot}$ with the stellar evolution code CLES \citep{cles}, adopting an overshooting of 0.2 pressure scale height and the solar chemical mixture of \citet{asplund}. We then compute the pulsation modes for $\ell$=1, 2 ($\ell_1, \ell_2$ hereafter), with the non-adiabatic code MAD including the implementation of the TAR as described in \citet{bouabid}. The periods in the inertial (observer) frame, $P_{in}$, are obtained from the periods in the corotating frame, $P_{co}$, through the following relation:

\begin{equation}
P_{in}=\frac{P_{co}}{1-m\frac{\eta}{2}},
\label{eq1}
\end{equation}
where $m$<0 corresponds to prograde modes. We present in Fig. \ref{figure1} the inertial periods of unstable modes throughout MS evolution of the 2.9 M$_{\odot}$ model for $\Omega/\Omega_{crit}$ from 0 to 0.80. We observe that the range of excited $P_{in}$ decreases to 0.2--0.4~d and becomes narrower as $\Omega$ increases. Our results are consistent with those obtained by T05: we find that the only excited modes along this track are prograde sectoral (PS), i.e. $m$=$-\ell$. They behave as Kelvin waves in which their $P_{co}$ first rise as rotation increases before converging to a constant value at higher rotation rates. Conversely, their $P_{in}$ will always decrease, due to the presence of $\eta$ in the denominator of Eq. \ref{eq1}.

Another consequence is the slight shift of the red border of the instability strip towards cooler temperatures (see also T05). Excitation is a balance between radiative damping, which depends on the radial order $k$, and driving. The efficiency of the latter requires the corotating period of the mode to be similar to the thermal relaxation timescale, $\tau_{\textrm{th}}$, in the driving region. The deeper this region is, the higher $\tau_{\textrm{th}}$ is. At lower $T_{\textrm{e}}$, the Fe opacity peak, which is the source of the driving, is deeper. Hence PS modes with a given order, $k$, will become unstable since their $P_{co}$ increases with rotation while their radiative damping does not change. Above a certain $\Omega$, the $P_{co}$ of PS modes no longer change, thereby limiting the number of excited PS modes. While T05 found unstable Kelvin waves at temperatures slightly lower than those we obtain ($\sim$0.03 dex in $\log T_{\textrm{e}}$), \citet{savonije} found none of them as unstable in a 3~M$_{\odot}$ model. These differences are certainly due to the stellar physics input used in the models, in particular the adopted chemical composition, on which depends the $\kappa$ mechanism. 

Table \ref{table1} reports the modes found to be unstable as $\Omega$ increases in the $\sim$20 Myr models, in accordance with the age of the \object{NGC 3766} cluster. We do not find excited modes in the 2.8~M$_{\odot}$ case. Thus, the limit of the instability band lies between 2.8 and 2.9~$M_{\odot}$. At that border, only $\ell_1$ PS modes are excited in the 2.9~M$_{\odot}$ model. As the stellar mass increases (3.2~M$_{\odot}$), both $\ell_1$ and $\ell_2$ PS are excited, with the $\ell_2$ showing $P_{in}$ smaller than those of $\ell_1$. The number of PS modes found to be unstable remains constant as rotation increases.

Some axisymmetric ($m=0$) modes are also unstable in the 3.2 M$_{\odot}$ case. Rotation decreases their $P_{in}$ down to 0.5 d. However, since their $P_{co}$ decrease as rotation increases, axisymmetric modes tend to stabilise and only a few of them remain unstable when rotation reaches $\Omega/\Omega_{\textrm{crit}}$=0.60 (see also T05). Finally most of retrograde modes ($m$>0) are stable since their $P_{co}$ are small and in addition undergo a severe decrease when rotation increases. So, the effect of rotation on the pulsation properties of stars defining the red border of the classic SPB instability strip, is to preferentially excite PS modes with low periods in the inertial frame. In particular, we note that these modes, in contrast to retrograde modes, are not expected to be affected by stabilising effects when rotational distortion of the star is taken into account \citep[][and ref. therein]{lee08}.

\section{Visibility of modes}
\label{section2}
To study the visibility of unstable modes described in Sect. \ref{section1}, let us introduce, for a separable system, the perturbation of a scalar eigenfunction for a given mode, $\xi_{\ell,m,\eta(k)}$, in terms of an infinite series of the usual spherical harmonics, $Y^m_l(\theta,\phi)$:

\begin{equation}
 \xi_{\ell,m,\eta}(r,\theta,\phi,t)= f(r) \ S \sum_{\ell} c_{\ell}(m,\eta) Y^m_{\ell}(\theta,\phi) \exp (i\omega t).
\label{eq2}
\end{equation}
In this expression, $f(r)$ expresses the radial dependence, $t$ is the time, $S$ is a constant such that $S \sum_l c_l Y^m_{\ell}(\theta,\phi)$ is the relative radial displacement at the photosphere, $\phi$ is the longitude and $c_{\ell}$ are the expansion coefficients. Within the TAR, the angular part of the eigenfunctions is actually given by the Hough functions, $\Theta^m_{l,\eta}$, which satisfy Laplace's tidal equation \citep[e.g.][]{leehough}. We simply expand $\Theta^m_{\ell,\eta}$ in terms of $Y^m_{\ell}$ to obtain the $c_{\ell}$ coefficients.    

\begin{figure}
\centering
\resizebox{\linewidth}{!}
{\includegraphics{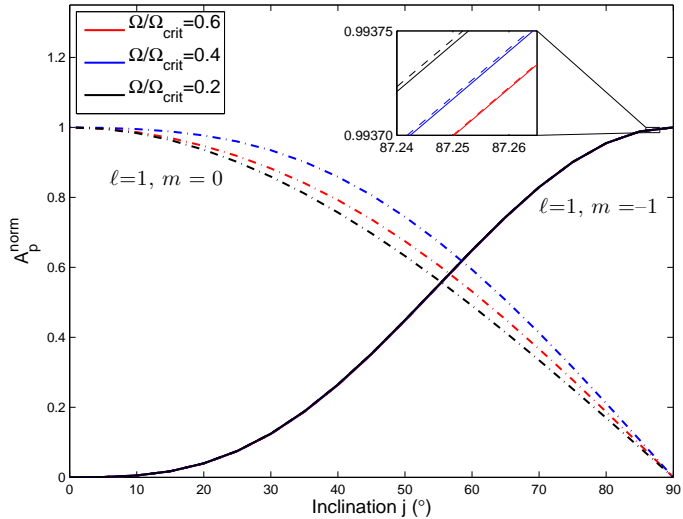}}
\caption{Normalised amplitudes, $A_p^{\textrm{norm}}$, as a function of the inclination. The continuous line represents the $\ell_1, k_{-16}, m_{-1}$ unstable mode of the 2.9 M$_{\odot}$ model. Dashed and dot-dashed lines show  the $\ell_1, k_{-16}, m_{-1}$ and $\ell_1, k_{-12}, m_0$ unstable modes of the 3.2 M$_{\odot}$ model, respectively. Both models are $\sim$20 Myr old. Black, blue and red colours correspond to rotations of 0.20, 0.40 and 0.60 $\Omega_{\textrm{crit}}$, respectively. The largest amplitudes, $\max A_p(\theta)$, reached by each mode are used for the normalisation.}
\label{figure2}
\end{figure}

We follow the approach of \citet{townsendvis} to determine in a photometric bandpass $p$, the magnitude variation ($\delta m_p$) caused by a non-radial pulsation in a rotating star:
\begin{eqnarray}
 &\delta m_p&=\ -\frac{2.5}{ln 10}S \sum_{\ell} c_{\ell} P^m_{\ell}(\cos j)b_{\ell} \bigg[ (1-\ell) (\ell+2) \cos(\omega t)  \nonumber \\ 
  &+& \left| \frac{\delta T_{\textrm{e}}}{T_{\textrm{e}}}\right| \frac{\partial \ln B_{\ell}}{\partial \ln T_{\textrm{e}}} \cos(\omega t + \psi_{T})+ \left| \frac{\delta g_e}{g_e}\right| \frac{\partial \ln B_{\ell}}{\partial \ln g_e} \cos(\omega t) \bigg],
\label{eq3}
\end{eqnarray}
where $P^m_{\ell}$ is an associated Legendre function, $j$ is the inclination angle, $\mid \delta T_{\textrm{e}}/T_{\textrm{e}}\mid$ and $\mid \delta g_e/g_e \mid$ are the variations of temperature and gravity at the photosphere, respectively, and $\psi_T$ is the phase lag between $\delta T_{\textrm{e}}/T_{\textrm{e}}$ and the radial displacement. Finally the terms $b_{\ell}=\int_{\lambda_1}^{\lambda_2}\int_0^1 h_{\lambda} \mu P_{\ell} d\mu d\lambda$ and $B_{\ell}=\int_{\lambda_1}^{\lambda_2}\int_0^1 F_{\lambda} h_{\lambda} \mu P_{\ell} d\mu d\lambda$ involve the limb darkening law ( $h_{\lambda}$) and the stellar outgoing monochromatic flux ($F_{\lambda}$), with $\mu=\cos \theta$ and $P_{\ell}$ a Legendre polynomial. The integration limits, $\lambda_1$--$\lambda_2$, correspond here to those of the CoRoT visible bandpass (courtesy of C. Barban and C. van't Veer-Menneret), as used in \citet{reesevis}.

\begin{figure*}[]
\centering
\includegraphics[width=0.5 \textwidth]{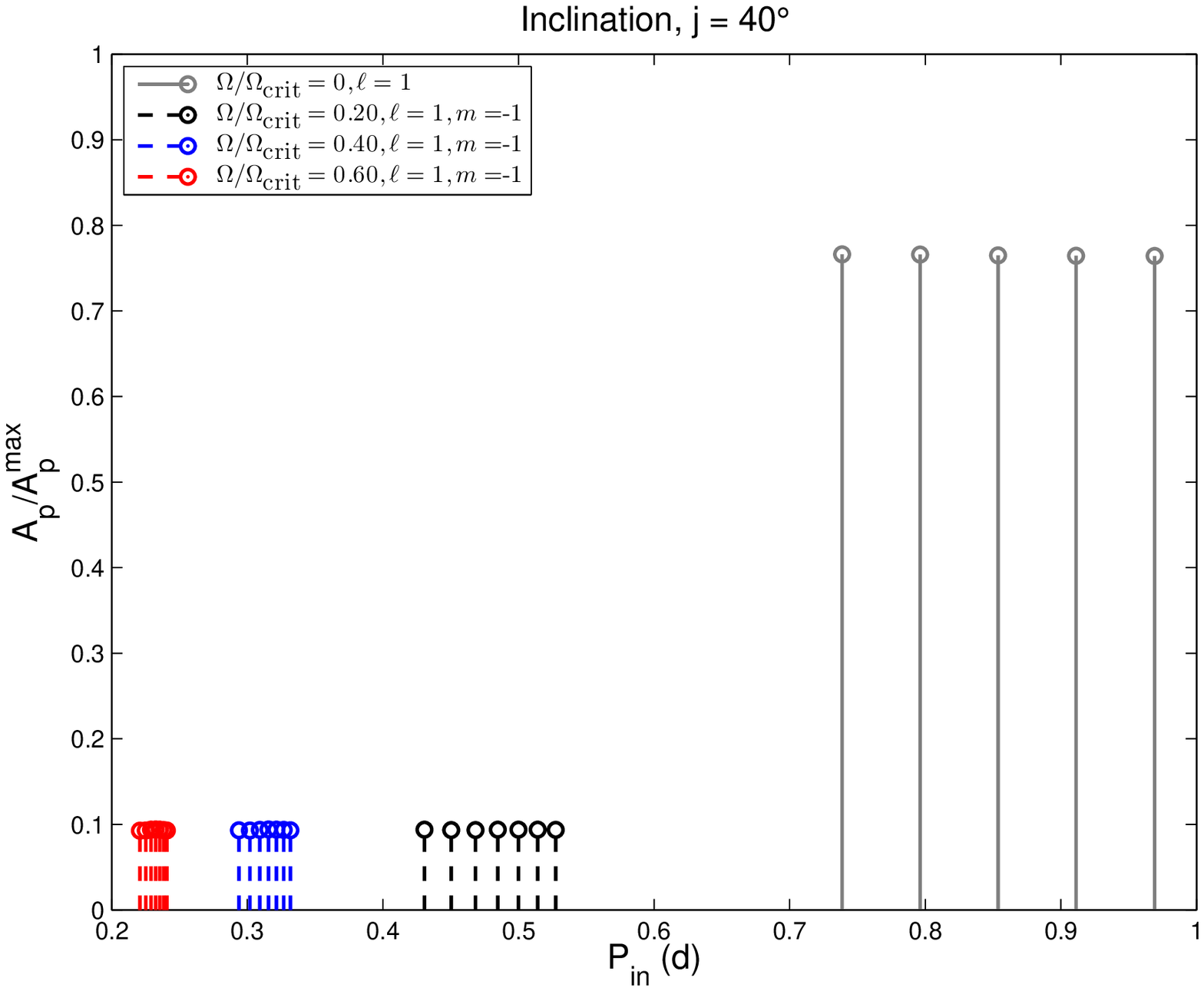}\includegraphics[width=0.5 \textwidth]{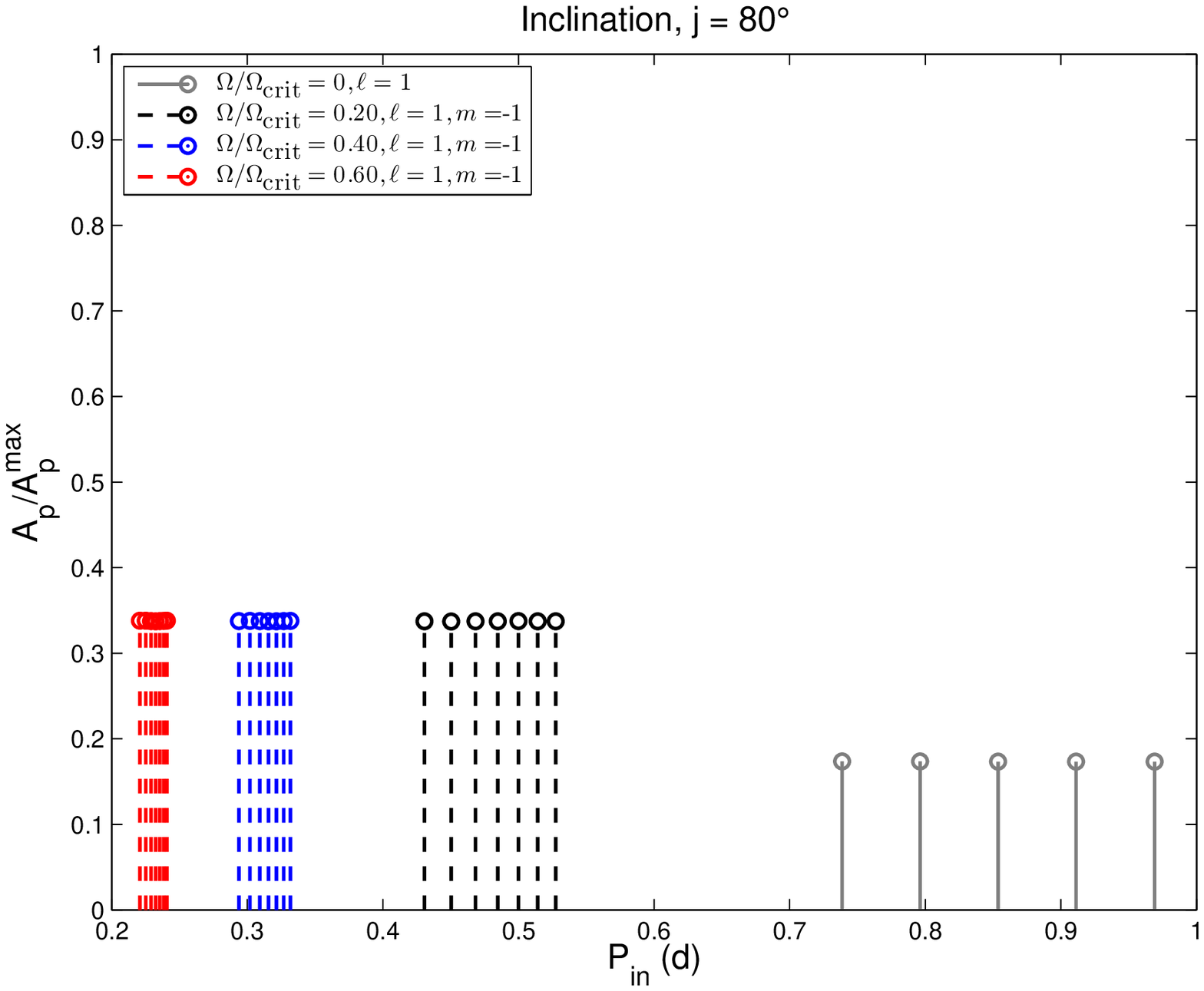}
\includegraphics[width=0.5 \textwidth]{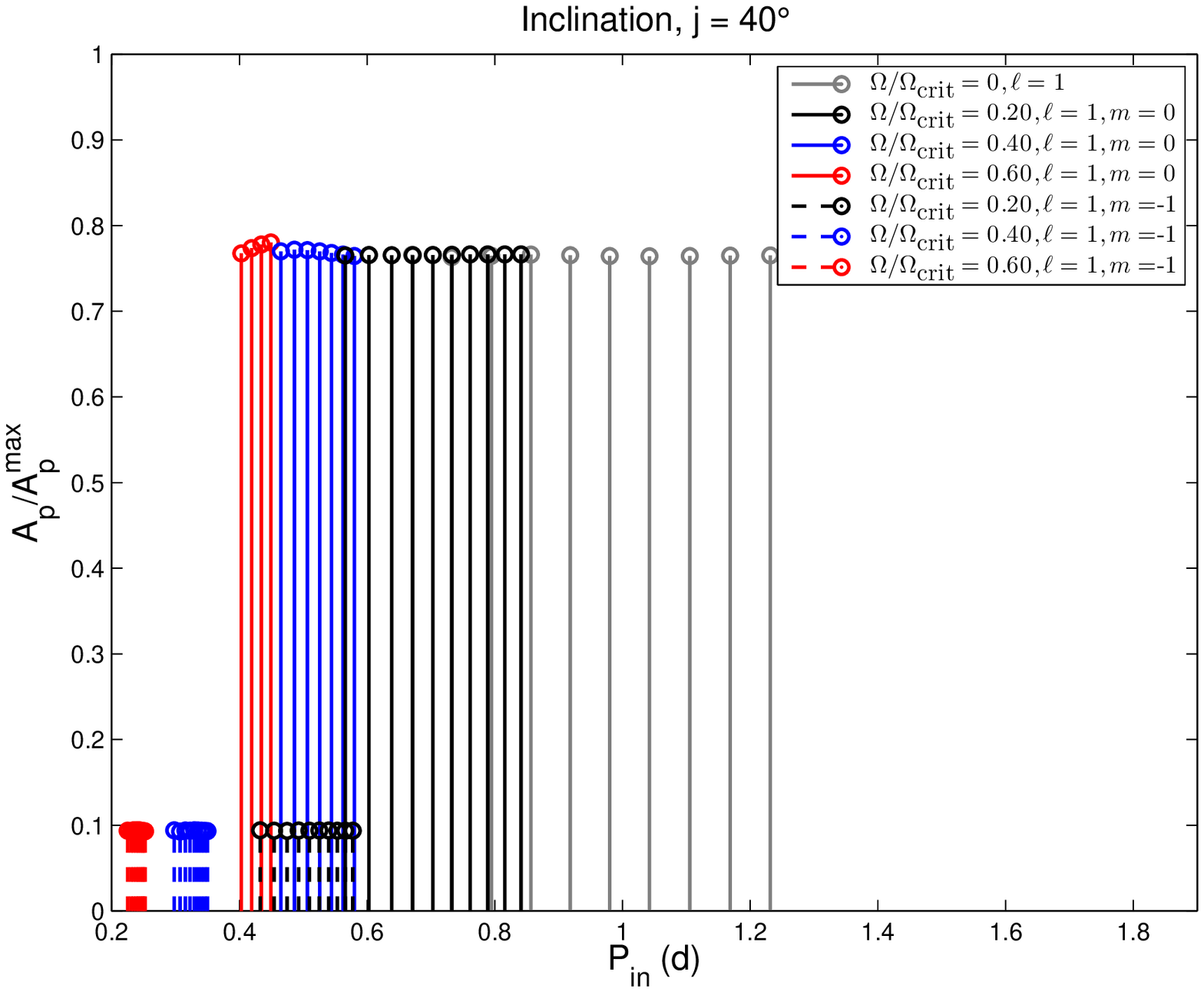}\includegraphics[width=0.5 \textwidth]{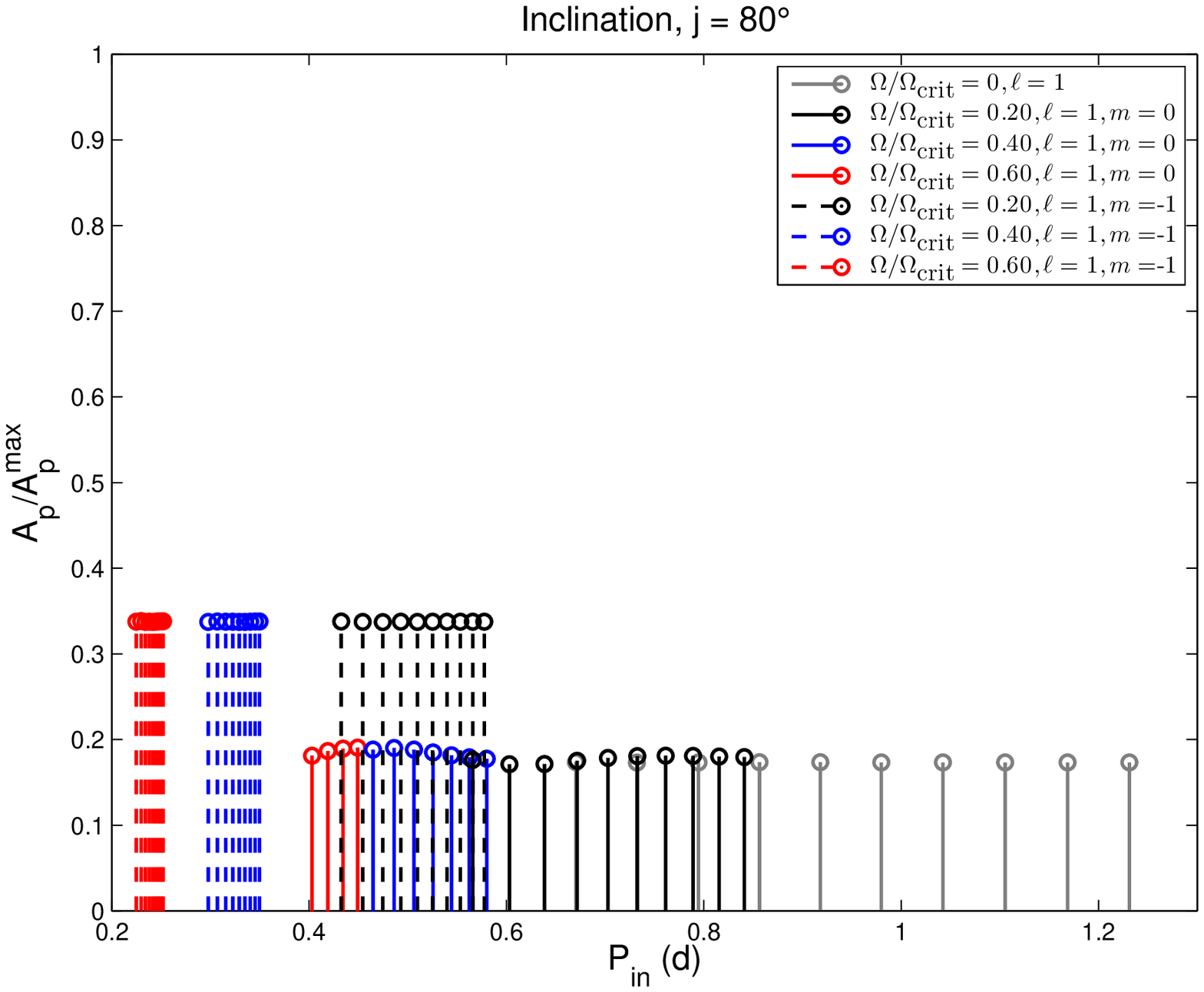}
\caption{Normalised amplitudes as a function of the inertial periods of all the $\ell_1$ modes found to be unstable in the 2.9 (top) and 3.2 (bottom) M$_{\odot}$ models aged $\sim$20 Myr. The detail of the normalisation are given in the main text. Left (resp. right) panels correspond to an observer viewing angle of 40$\degr$ (resp. 80 $\degr$). Grey solid lines show the modes in a case with no rotation. Dashed and solid lines represent the PS and axisymmetric modes, respectively, while black, blue and red colours refer to a rotation of 0.20, 0.40 and 0.60$\Omega_{\textrm{crit}}$, respectively.}
\label{figure2bis}
\end{figure*}

The amplitudes of the perturbation ($\delta m_p$), which we refer to as $A_p$, are represented in Fig.~\ref{figure2} for three modes: the $\ell_1, k_{-16}, m_{-1}$ mode of the 2.9 M$_{\odot}$ model and the $\ell_1, k_{-16}, m_{-1}$ and $\ell_1, k_{-12}, m_0$ modes of the 3.2 M$_{\odot}$ model. They are shown for rotation rates of 0.20, 0.40 and 0.60 $\Omega_{\textrm{crit}}$, these modes being excited in every case. The $A_p$ are normalised by their maximum value.

Since they correspond to Kelvin waves, the $\ell_1$ PS modes appear as expected, clearly confined to the equator \citep[see e.g.][for details on the nature of Kelvin waves]{gill}. In our case, the global shape of the Hough functions of PS modes varies slightly as $\Omega$ (and so $\eta$) increases \citep[see also][]{dd}. Thus the normalised PS visibilities are insensitive to $\Omega$ and overlap each other in Fig.~\ref{figure2}. The $\ell_2$ PS modes (not shown) are similarly confined to the equator.

On the contrary, the visibilities of $\ell_1$ axisymmetric modes are maximum when the star is seen towards the pole. The geometry of the $\ell_1, m_0$ Hough functions is also at the origin of this trend. We do not consider the other prograde and retrograde modes of the 3.2 M$_{\odot}$ model, since they are no longer unstable when rotation increases.

Furthermore, we compare the visibilities of the PS and axisymmetric modes with those of modes in a non-rotating case. Determining the exact amplitudes of modes actually requires including non-linear processes, which is still a theoretical issue. When the eigenfunctions are normalised to the photospheric relative radial displacement, the predicted relative variations of effective temperature can be very different between modes of a given $\ell$ and $m$ series. Indeed, $\delta T_{\textrm{e}}/T_{\textrm{e}}$ scales as $\lambda/\omega_{\textrm{co}}^2$ for gravito-inertial modes with such a normalisation. We can however expect that the saturation mechanism actually limiting the growth of mode amplitudes in real stars, does not favour any particular value of $|\delta T_{\textrm{e}}/T_{\textrm{e}}|$. We thus adopt $S\propto |\delta T_{\textrm{e}}/T_{\textrm{e}}|^{-1}$ in Eq.~(\ref{eq3}) to avoid any non-physical dependence. 

In order to easily compare the modes between themselves, we now normalise them by a unique value corresponding to the largest amplitude $A_p(\theta)$ reached by one of them, as suggested in \citet{savonije2}. We show in Fig.~\ref{figure2bis} the amplitudes normalised in this way for all the $\ell_1$ modes found to be excited in the two models considered so far. For each model, the normalising factor corresponds to the amplitude of the non-rotating $\ell_1, m_0$ mode with the highest period, seen pole-on. For an inclination $j$=40$\degr$, the PS mode visibilities are lower than those of the non-rotating modes by a factor of $\sim$8. The amplitudes of axisymmetric modes, that are only excited in the 3.2 M$_{\odot}$ model, are of same order as the non-rotating ones, although the former see their periods shortened by rotation. As expected from Fig.~\ref{figure2}, the visibility pattern of the PS modes is not altered when rotation is increased. 

Considering $j$=80$\degr$, the PS modes of both models present visibilities higher by up to a factor of 2 than those of their respective non-rotating modes. The same is true for the comparison with the axisymmetric modes in the more massive model. We do not illustrate the case of $\ell_2$ modes in Fig.~\ref{figure2bis}, since their normalised amplitudes never exceed 0.1, regardless of the $\Omega$ and $j$ values.

Hence when the star is seen towards the equator, we can expect the $\ell_1$ PS modes to display amplitudes of $\sim$30$-$40\% of the largest amplitude shown by a non-rotating SPB mode. This visibility ratio is of the same order as the one observed in Mo13 between the amplitudes of the SPBs and new variable stars. The low-degree PS modes of fast rotating stars seen nearly equator-on thus seem to be good candidates for explaining the origin of this new variability: they both present the correct amplitudes and a shift of their periods to the 0.2$-$0.5 d range and they are the most numerous found to be unstable.

\section{Gravity darkening due to rotation}

We now inspect the effects of rotation and inclination on visual properties of the stars themselves, to determine whether SPB pulsators might be observed outside their predicted domain in the HR diagram. We still consider solid-body rotation. The layers of a rotating star are defined by isobar contours that, for solid-body rotation, coincide with equipotentials. Interestingly, the effect of distortion on the inner layers depends on the mass of the star. SPB stars are among the least affected, maintaining a ratio <1\% between the polar radius at $\Omega$=0 and the polar radii of models up to 0.80 $\Omega_{\textrm{crit}}$ \citep[see Fig. 2 of][their 3 M$_{\odot}$ track]{ekstrom}. Their stellar surface can then be described very simply by a Roche model:

\begin{equation}
 \frac{GM}{R(\theta)}+\frac{1}{2}\Omega^2R^2(\theta)\sin^2\theta=\frac{GM}{R_{\textrm{pol}}},
\label{eq4}
\end{equation}
where $R$ is the radius of the surface at colatitude $\theta$ and $R_{\textrm{pol}}$ the polar radius. Consequently the effective gravity, $g_e$, depends only on $R$ and $\Omega$ \citep[see its expression in e.g.][]{maederformula}, and using von Zeipel's theorem, we express $T_{\textrm{e}}$ as a function of $g_e$:

\begin{equation}
 T_{\textrm{e}}(\theta)=\left[\frac{L \bar{\rho}}{2M\sigma(2\pi G \bar{\rho}-\Omega^2)}\right]^{\frac{1}{4}}g_e^{\frac{1}{4}}(\theta),
\label{eq5}
 \end{equation}
where $\bar{\rho}$ is the mean stellar density and $\sigma$ is the Stefan-Boltzmann constant. The specific intensity over the whole electromagnetic spectrum is provided by $I=\sigma T_{\textrm{e}}^4/\pi$ . After integrating the specific intensity and Eq. \ref{eq5} over the projected stellar surface (with help of Eq. \ref{eq4}) as seen by an observer, we obtain the apparent luminosity and effective temperature respectively. 

We use solid-body rotating models obtained with the Geneva code \citep{genevacode} to determine the impact of gravity darkening in the 2.9 and 3.2 M$_{\odot}$ cases. We make the computations for rotation rates from 0.10 to 0.99 $\Omega_{\textrm{crit}}$. Following the above prescription, we derive $T_{\textrm{e}}$ and $L$ for various inclinations. The effects of limb darkening are included with help of the bolometric coefficients derived by \citet{howarth} for this purpose. We also compare non-rotating Geneva models with CLES models and find that for a given mass they coincide in the HR diagram, ensuring that the Geneva models can be consistently used for our analysis.
%

\begin{figure}
\centering
\resizebox{ \linewidth}{!}
{\includegraphics{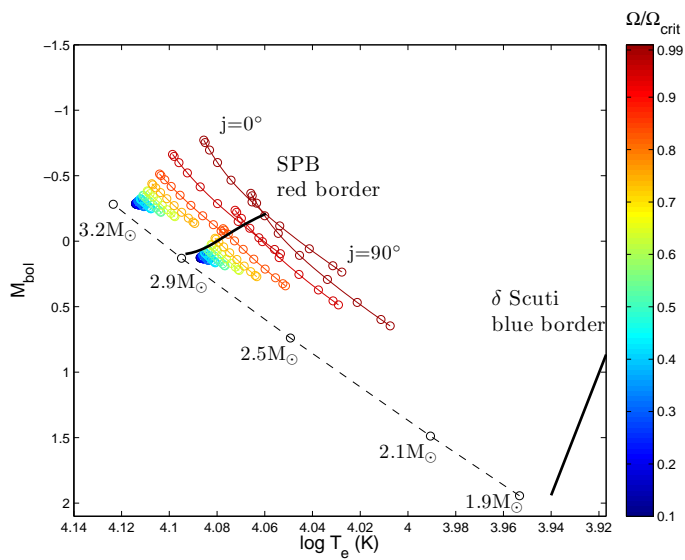}}
\caption{Bolometric magnitude and effective temperature as a function of $j$, for 2.9 and 3.2 M$_{\odot}$ Geneva models $\sim$20 Myr old. Their rotation rate is indicated by the colour scale. The $j$ angles from 0 to 90$\degr$ are shown with help of dots every 10$\degr$. The dot-dashed line represents the zero age MS of non-rotating models from 1.9 to 3.2 M$_{\odot}$. Thick black lines show the red and blue borders of the instability strips for non-rotating SPB (this work) and $\delta$ Scuti \citep{dupretscuti} models, respectively.} 
\label{figure3}
\end{figure}

The Geneva models are shown in Fig.~\ref{figure3} for an isochrone of $\sim$20 Myr, where $L$ has been previously converted to bolometric magnitude, $M_{\textrm{bol}}$. Those models span over a bit less than 2 magnitudes if one considers a rotation rate of 0.99 $\Omega_{\textrm{crit}}$. With the help of models of the same mass but computed with CLES, we have already shown that the PS modes are the modes most likely to be excited for $\Omega/\Omega_{\textrm{crit}}\geq0.40$. We have also determined in Fig.~\ref{figure2bis} that the $\ell_1$ modes are the most visible in the oscillation spectra precisely for the angles under which the Geneva models are the most affected by gravity darkening, thereby appearing considerably cooler and fainter. 

Moreover, we have placed in Fig.~\ref{figure3} the red and blue borders of the ``classic'' non-rotating SPB and $\delta$ Scuti instability strips, respectively, obtained from the stability analysis of CLES models. The Geneva models, selected to be representative of rotating SPB stars at the red border of the strip, appear as if they were located in the theoretical gap between SPB and $\delta$ Scuti stars. Therefore gravity darkening is able to displace fast rotating SPB stars well into this gap, meanwhile the corresponding effect of the Coriolis force shifts downwards the pulsation period domain as well as the amplitudes of excited oscillation modes. Unfortunately Mo13 used an uncalibrated photometric system, thereby hindering a direct comparison with our results.      

\section{Discussion}

The PS modes appear to be those dominantly excited in the red border of the SPB instability strip. Owing to their geometry, these modes are also the most visible oscillations when the star is viewed nearly equator-on, which corresponds to the situation where stars are the most affected by gravity darkening. We now would like to assess whether we are able to explain, at least qualitatively, the other observational properties reported for NGC~3766. \citet{mowlavi2} gave the fraction of periodic variables by range of magnitude for this cluster. While $\sim$40-50\% of the stars in the SPB domain are periodic variables, this fraction monotonically decreases to $\sim$10\% in the gap between SPB and $\delta$ Scuti stars. This trend can be explained due to the combined effect of the shift of the fast rotating SPBs towards cooler regions and the increase, according to the stellar mass function, of less massive (non pulsating) stars populating this HR region.

Mo13 also found, close to the transition with their SPB candidates (periods >1 d), multi-periodic new variable stars (periods <1 d), including one hybrid with periods both smaller and larger than 1~d. We have therefore also analysed (not presented here) models of 3.5 and 4 M$_{\odot}$ rotating at $\Omega/\Omega_{crit}$=0.40 and 0.60. As a general result, about 90 and 80\% of up to a hundred excited modes in the respective 3.5 and 4 M$_{\odot}$ models have periods smaller than 1~d. Hence, the objects near the transition could correspond to more massive fast rotating SPBs. On the other hand, the Mo13 SPB candidates belong to a brighter sample of stars. \citet{decatrev} showed that the domain of pulsations in SPBs with $v \sin i$<100 km/s is typically composed of periods larger than 1~d. We suggest that the Mo13 brighter sample of stars could correspond to slowly rotating SPBs: given their low rotation rate, no significant shift is expected with respect to the HR location of the classic SPB instability strip.

We notice that \citet{savonije2} advanced Rossby and Yanai waves as an explanation to the presence of low-period modes in late B-type stars. He analysed models of 3 to 8 M$_{\odot}$ in different evolutionary stages, though none of them were as young as the models of the present work. In particular, he pointed out that the Rossby waves in the less massive of his models were only excited at late stages. We thus check for the stability of Yanai modes, also referred to in the literature as retrograde mixed modes \citep{townsendretro} or mixed Rossby-Poincar\'{e} modes. Following the classification of \citet{leehough}, a quantum number\footnote{not to be mixed up with the radial order for which we have referred to by also using $k$ in this work} $k=-1$ can be attributed to these modes ($k\leq-2$ for Rossby waves). Their displacement is largely dominated by the toroidal component \citep[see e.g.][]{saiormode}. As a consequence, they do not undergo equatorial confinement, as opposed to the PS modes. Note also that Yanai modes present a hybrid character since their restoring force results from both the Coriolis force and buoyancy.


We therefore look at the stability of the $k=-1$, $m_1$ and $m_2$ modes in the 2.9 and 3.2 M$_{\odot}$ ($\sim$20 Myr) models  for rotation rates of 0.20, 0.40 and 0.60 $\Omega_{\textrm{crit}}$. The results, presented in Table~\ref{table2}, are fully consistent with those of \citet{savonije} and \citet{townsendretro}. In the 2.9 M$_{\odot}$ model, $m_1$ modes are only found to be unstable at the lowest rotation rate, with periods of several days. The $m_2$ modes are only excited for $\Omega/\Omega_{\textrm{crit}}$= 0.20 and 0.40 and concentrate in very narrow ranges of period below 1 d. The situation is similar in the more massive model, except that $m_1$ and $m_2$ modes are excited for each of the rotation rates we have looked at. These results are in good agreement with those presented by the above authors.

\begin{table}[!]
\caption{Same as Table~\ref{table1}, but now for the instability domains of Yanai modes. The equatorial velocities are the same as those in Table~\ref{table1} and are no longer indicated.}             
\label{table2}      
\centering          
\begin{tabular}{l l c c}     
\hline\hline       
Mass & $\Omega/\Omega_{\textrm{crit}}$ & $m=1$ & $m=2$   \\ 
\hline  
2.9 & 0.20 & 5.09 - 8.63 (6) & 0.625 - 0.640 (8)  \\ 
 & 0.40 & -- & 0.298 - 0.304 (5)   \\ 
 & 0.60 & -- & --   \\ 
3.2 & 0.20 & 4.05 - 9.57 (11)  & 0.636 - 0.659 (10) \\ 
 & 0.40 & 1.16 - 1.84 (9) & 0.298 - 0.316 (11) \\ 
 & 0.60 & 0.64 - 0.81 (7) & 0.192 - 0.203 (9) \\ 
\hline                  
\end{tabular}
\end{table}

As we have already mentioned, the Yanai modes have a rather different geometry than PS (Kelvin) modes. To test whether they could also correspond to the variability detected by Mo13, we compare the visibility of the Yanai modes with that of non-rotating modes, applying the same procedure as in Sect.~\ref{section2}. Since these modes present no confinement to the equator \citep[see also][]{savonije2}, they reach their maximum of visibility for a stellar inclination around 45$\degr$ and they are almost unseen for $j$=10$\degr$ and 80$\degr$. In both models, the maximum amplitudes of $m_2$ modes never exceed $\sim$10\% of those of non-rotating SPB modes. The amplitudes of $m_1$ modes reach a maximum of $\sim$50\% when the star has an inclination around $j$=45$\degr$ while this value drops to $\sim$20\% for $j$=10$\degr$ and $\sim$10\% for $j$=80$\degr$. These percentages are still expressed in terms of the amplitudes of non-rotating SPB modes, and are valid regardless of the rotation rate. Since only $m_1$ modes might be observable, furthermore for inclinations close to $j$=45$\degr$, we do not expect them to be detected in stars significantly displaced outside the SPB instability strip by gravity darkening. Moreover the $m_1$ Yanai modes would present periods close to those of non-rotating SPB stars. We thus think that the few Mo13 new variables with periods larger than 0.7~d could correspond to either Yanai $m_1$ modes or axisymmetric and retrograde classic g modes, expected to be rare.

We finally look at the case of the NGC~884 cluster, which is also known to host variable stars similar to those detected by Mo13. Among the objects of this cluster for which fundamental parameters were determined \citep{saesen2}, one star, identified as Oo~2151, is clearly outside the SPB band with $\log T_{\textrm{e}}<4$, though this value is declared as uncertain. Similarly to the situation in NGC~3766, its amplitude in the V (Geneva) bandpass of 2 mmag is among the lowest of NGC~884 pulsators (including SPBs). Our results support this feature as well as the period of $\sim$0.18 d detected for this star. Determining the rotational velocity of Oo~2151 and its fundamental parameters with a better accuracy, would help to clarify its status. 

\citet{degrootecorot} reported the detected variability of stars in the field of view of the initial observing run (IRa01) of the CoRoT space mission. For stars with a periodic variability, they classified, based on the main frequencies, 43 targets as $\beta$~Cephei stars. However, a refined analysis revealed that only one of them was a good $\beta$~Cephei candidate. Instead, the atmospheric parameters ($T_{\textrm{e}}$ and $\log g$) of these IRa01 targets  estimated from Str\"omgren photometry (u,b,v filters), located them along the SPB instability strip as well as between the SPB and $\delta$~Scuti instability strips, in much the same way as in Mo13. Moreover, \citet{degrootecorot} distinguished at least two different kinds of frequency spectra among these ``new-variables'': those with dense spectra dominated by frequencies in the $\beta$~Cephei domain (P$\sim$0.1--0.3 d) but also containing low-frequency signatures, all of them of low amplitude; and those presenting more dominant frequencies (still in the range of 0.1--0.3 d) of slightly larger amplitudes. The first group is found in the region close to the red border of the classical SPB instability strip, and the second one in the coldest border of this new observational instability strip. The properties of the first group could correspond to the one we refer to as the transition region in the Mo13 case, with pulsation periods larger and lower than 1~d, and hence likely pulsate with low~amplitude Yanai or classic g modes. The second group could match the properties of PS modes in rapid rotators seen equator-on, with relatively larger amplitudes (see Sect.~\ref{section2}). These stars would also require spectroscopic campaigns to proceed to a better determination of their characteristics.

\section{Conclusion}

In this paper, we have analysed a scenario able to explain the recent discovery of ``new variable'' stars in NGC~3766. Since this cluster is known to host a large number of fast rotators, it is relevant to include the effects of rotation in the study of the seismic properties of its members. In that aim, we have used the TAR and showed that PS modes of fast rotating SPB stars are good candidates to explain this new kind of variability. Indeed, in models at the red border of the classic SPB instability strip, the period domain of excited modes decreases from 0.7--1.2~d (no rotation) to 0.15--0.6~d or 0.12--0.45~d depending on the rotation rate. In particular, the domain of PS modes decreases to 0.12--0.35~d. The number of excited PS modes remains stable while they are confined towards the equator as rotation increases, due to their Kelvin wave nature. This makes them only visible in highly inclined stars, i.e. stars with low latitudes facing the observer. 

Meanwhile, these conditions of fast rotation and high inclination also lead to a maximum shift of the effective temperature and luminosity downwards, displacing these low-period, low-amplitude pulsators outside the classic instability strip of SPBs, in the gap between this latter strip and that of $\delta$ Scuti stars. Furthermore, for such inclinations, the visibilities of $\ell_1$ PS modes in rotating SPBs are qualitatively 2 to 3 times lower than in the non-rotating stars. These values reproduce the ratio of the variability amplitudes observed by Mo13 between the stars classified as SPB and ``new variable'', respectively.

Other modes excited by rotation, mainly the Yanai modes, cannot satisfactorily reproduce all the properties of Mo13 variables. In particular, we have found that the Yanai with the lowest periods ($\lesssim$0.65~d) are also those with the lowest visibility, clearly lower than that of PS modes.

Fast rotating SPB stars could also explain the similar variable stars observed by CoRoT \citep{degrootecorot} and in the \object{NGC 884} cluster \citep{saesenpersei}.

The domain of validity of the TAR remains uncertain at high rotation rates. A new analysis with non-adiabatic two dimensional treatment (\citealp{espinosa}, Reese et al., in prep) would lead to more quantitative predictions, although the computation of low frequency gravity modes remains a real numerical challenge.

\begin{acknowledgements}
     We are grateful to A. Noels, C. Georgy, S. Ekstr\"{o}m and A. Granada for helpful discussions. SJAJS was supported by a F.R.I.A. fellowship from the National Fund for Scientific Research, Belgium. DRR thanks the support of ``Subside f\'ed\'eral de la recherche 2012'',  Universit\'e de Li\`{e}ge, and is currently funded by the European Community's Seventh Framework Programme (FP7/2007-2013) under grant agreement no. 312844 (SPACEINN).
\end{acknowledgements}
\bibliographystyle{aa}
\bibliography{biblio}

\newcommand{\noopsort}[1]{}
\begin{thebibliography}{38}
\expandafter\ifx\csname natexlab\endcsname\relax\def\natexlab#1{#1}\fi

\bibitem[{{Aidelman} {et~al.}(2012){Aidelman}, {Cidale}, {Zorec}, \&
  {Arias}}]{aidelman}
{Aidelman}, Y., {Cidale}, L.~S., {Zorec}, J., \& {Arias}, M.~L. 2012, \aap,
  544, A64

\bibitem[{{Asplund} {et~al.}(2005){Asplund}, {Grevesse}, \& {Sauval}}]{asplund}
{Asplund}, M., {Grevesse}, N., \& {Sauval}, A.~J. 2005, in ASP Conf. Ser., Vol.
  336, Cosmic Abundances as Records of Stellar Evolution and Nucleosynthesis,
  ed. {T.~G.~Barnes III \& F.~N.~Bash}, 25

\bibitem[{{Ballot} {et~al.}(2012){Ballot}, {Ligni{\`e}res}, {Prat}, {Reese}, \&
  {Rieutord}}]{ballottar}
{Ballot}, J., {Ligni{\`e}res}, F., {Prat}, V., {Reese}, D.~R., \& {Rieutord},
  M. 2012, in ASP Conf. Ser., Vol. 462, Progress in Solar/Stellar Physics with
  Helio- and Asteroseismology, ed. H.~{Shibahashi}, M.~{Takata}, \& A.~E.
  {Lynas-Gray}, 389

\bibitem[{{Berthomieu} {et~al.}(1978){Berthomieu}, {Gonczi}, {Graff},
  {Provost}, \& {Rocca}}]{berthomieu}
{Berthomieu}, G., {Gonczi}, G., {Graff}, P., {Provost}, J., \& {Rocca}, A.
  1978, \aap, 70, 597

\bibitem[{{Bouabid} {et~al.}(2013){Bouabid}, {Dupret}, {Salmon},
  {Montalb{\'a}n}, {Miglio}, \& {Noels}}]{bouabid}
{Bouabid}, M.-P., {Dupret}, M.-A., {Salmon}, S., {et~al.} 2013, \mnras, 429,
  2500

\bibitem[{{Daszynska-Daszkiewicz} {et~al.}(2007){Daszynska-Daszkiewicz},
  {Dziembowski}, \& {Pamyatnykh}}]{dd}
{Daszynska-Daszkiewicz}, J., {Dziembowski}, W.~A., \& {Pamyatnykh}, A.~A. 2007,
  \actaa, 57, 11

\bibitem[{{De Cat}(2002)}]{decatrev}
{De Cat}, P. 2002, in Astronomical Society of the Pacific Conference Series,
  Vol. 259, IAU Colloq. 185: Radial and Nonradial Pulsationsn as Probes of
  Stellar Physics, ed. C.~{Aerts}, T.~R. {Bedding}, \&
  J.~{Christensen-Dalsgaard}, 196

\bibitem[{{De Cat} {et~al.}(2007){De Cat}, {Briquet}, {Aerts}, {Goossens},
  {Saesen}, {Cuypers}, {Yakut}, {Scuflaire}, {Dupret}, {Uytterhoeven}, {van
  Winckel}, {Raskin}, {Davignon}, {Le Guillou}, {van Malderen}, {Reyniers},
  {Acke}, {De Meester}, {Vanautgaerden}, {Vandenbussche}, {Verhoelst},
  {Waelkens}, {Deroo}, {Reyniers}, {Ausseloos}, {Broeders},
  {Daszy{\'n}ska-Daszkiewicz}, {Debosscher}, {De Ruyter}, {Lefever}, {Decin},
  {Kolenberg}, {Mazumdar}, {van Kerckhoven}, {De Ridder}, {Drummond}, {Barban},
  {Vanhollebeke}, {Maas}, \& {Decin}}]{decatmaia}
{De Cat}, P., {Briquet}, M., {Aerts}, C., {et~al.} 2007, \aap, 463, 243

\bibitem[{{Degroote} {et~al.}(2009{\natexlab{a}}){Degroote}, {Aerts},
  {Ollivier}, {Miglio}, {Debosscher}, {Cuypers}, {Briquet}, {Montalb{\'a}n},
  {Thoul}, {Noels}, {De Cat}, {Balaguer-N{\'u}{\~n}ez}, {Maceroni}, {Ribas},
  {Auvergne}, {Baglin}, {Deleuil}, {Weiss}, {Jorda}, {Baudin}, \&
  {Samadi}}]{degrootecorot}
{Degroote}, P., {Aerts}, C., {Ollivier}, M., {et~al.} 2009{\natexlab{a}}, \aap,
  506, 471

\bibitem[{{Degroote} {et~al.}(2009{\natexlab{b}}){Degroote}, {Miglio},
  {Debosscher}, {Montalb{\'a}n}, {Cuypers}, {Briquet}, {De Cat}, {Thoul},
  {Morel}, {Niemczura}, {Balaguer-N{\'u}{\~n}ez}, {Maceroni}, {Ribas}, {Noels},
  {Aerts}, {Auvergne}, {Baglin}, {Catala}, {Deleuil}, {Michel}, {Ollivier},
  {Jorda}, \& {Samadi}}]{degroote09coast}
{Degroote}, P., {Miglio}, A., {Debosscher}, J., {et~al.} 2009{\natexlab{b}},
  Communications in Asteroseismology, 158, 167

\bibitem[{{Dupret} {et~al.}(2005){Dupret}, {Grigahc{\`e}ne}, {Garrido},
  {Gabriel}, \& {Scuflaire}}]{dupretscuti}
{Dupret}, M.-A., {Grigahc{\`e}ne}, A., {Garrido}, R., {Gabriel}, M., \&
  {Scuflaire}, R. 2005, \aap, 435, 927

\bibitem[{{Eckart}(1963)}]{eckart}
{Eckart}, C. 1963, Quarterly Journal of the Royal Meteorological Society, 89,
  567

\bibitem[{{Eggenberger} {et~al.}(2008){Eggenberger}, {Meynet}, {Maeder},
  {Hirschi}, {Charbonnel}, {Talon}, \& {Ekstr{\"o}m}}]{genevacode}
{Eggenberger}, P., {Meynet}, G., {Maeder}, A., {et~al.} 2008, \apss, 316, 43

\bibitem[{{Ekstr{\"o}m} {et~al.}(2008){Ekstr{\"o}m}, {Meynet}, {Maeder}, \&
  {Barblan}}]{ekstrom}
{Ekstr{\"o}m}, S., {Meynet}, G., {Maeder}, A., \& {Barblan}, F. 2008, \aap,
  478, 467

\bibitem[{{Espinosa Lara} \& {Rieutord}(2013)}]{espinosa}
{Espinosa Lara}, F. \& {Rieutord}, M. 2013, \aap, 552, A35

\bibitem[{Gill(1982)}]{gill}
Gill, A. 1982, Atmosphere-Ocean Dynamics (Academic press, London)

\bibitem[{{Howarth}(2011)}]{howarth}
{Howarth}, I.~D. 2011, \mnras, 413, 1515

\bibitem[{{Lee}(2008)}]{lee08}
{Lee}, U. 2008, Communications in Asteroseismology, 157, 203

\bibitem[{{Lee} \& {Saio}(1997)}]{leehough}
{Lee}, U. \& {Saio}, H. 1997, \apj, 491, 839

\bibitem[{{Maeder} \& {Meynet}(2012)}]{maederformula}
{Maeder}, A. \& {Meynet}, G. 2012, Reviews of Modern Physics, 84, 25

\bibitem[{{Maeder} \& {Peytremann}(1972)}]{maederpeyt}
{Maeder}, A. \& {Peytremann}, E. 1972, \aap, 21, 279

\bibitem[{{McSwain} {et~al.}(2008){McSwain}, {Huang}, {Gies}, {Grundstrom}, \&
  {Townsend}}]{mcswain}
{McSwain}, M.~V., {Huang}, W., {Gies}, D.~R., {Grundstrom}, E.~D., \&
  {Townsend}, R.~H.~D. 2008, \apj, 672, 590

\bibitem[{{Mowlavi} {et~al.}(2013{\natexlab{a}}){Mowlavi}, {Barblan}, {Saesen},
  \& {Eyer}}]{mowlavi}
{Mowlavi}, N., {Barblan}, F., {Saesen}, S., \& {Eyer}, L. 2013{\natexlab{a}},
  \aap, 554, A108

\bibitem[{{Mowlavi} {et~al.}(2013{\natexlab{b}}){Mowlavi}, {Saesen}, {Barblan},
  \& {Eyer}}]{mowlavi2}
{Mowlavi}, N., {Saesen}, S., {Barblan}, F., \& {Eyer}, L. 2013{\natexlab{b}},
  ArXiv e-prints

\bibitem[{{Pamyatnykh}(1999)}]{pami99}
{Pamyatnykh}, A.~A. 1999, \actaa, 49, 119

\bibitem[{{Reese} {et~al.}(2013){Reese}, {Prat}, {Barban}, {van 't
  Veer-Menneret}, \& {MacGregor}}]{reesevis}
{Reese}, D.~R., {Prat}, V., {Barban}, C., {van 't Veer-Menneret}, C., \&
  {MacGregor}, K.~B. 2013, \aap, 550, A77

\bibitem[{{Saesen} {et~al.}(2013){Saesen}, {Briquet}, {Aerts}, {Miglio}, \&
  {Carrier}}]{saesen2}
{Saesen}, S., {Briquet}, M., {Aerts}, C., {Miglio}, A., \& {Carrier}, F. 2013,
  \aj, 146, 102

\bibitem[{{Saesen} {et~al.}(2010){Saesen}, {Carrier}, {Pigulski}, {Aerts},
  {Handler}, {Narwid}, {Fu}, {Zhang}, {Jiang}, {Vanautgaerden}, {Kopacki},
  {St{\c e}{\'s}licki}, {Acke}, {Poretti}, {Uytterhoeven}, {Gielen},
  {{\O}stensen}, {De Meester}, {Reed}, {Ko{\l}aczkowski}, {Michalska},
  {Schmidt}, {Yakut}, {Leitner}, {Kalomeni}, {Cherix}, {Spano}, {Prins}, {van
  Helshoecht}, {Zima}, {Huygen}, {Vandenbussche}, {Lenz}, {Ladjal}, {Puga
  Antol{\'{\i}}n}, {Verhoelst}, {De Ridder}, {Niarchos}, {Liakos}, {Lorenz},
  {Dehaes}, {Reyniers}, {Davignon}, {Kim}, {Kim}, {Lee}, {Lee}, {Kwon},
  {Broeders}, {van Winckel}, {Vanhollebeke}, {Waelkens}, {Raskin}, {Blom},
  {Eggen}, {Degroote}, {Beck}, {Puschnig}, {Schmitzberger}, {Gelven},
  {Steininger}, {Blommaert}, {Drummond}, {Briquet}, \&
  {Debosscher}}]{saesenpersei}
{Saesen}, S., {Carrier}, F., {Pigulski}, A., {et~al.} 2010, \aap, 515, A16

\bibitem[{{Saio}(1982)}]{saiormode}
{Saio}, H. 1982, \apj, 256, 717

\bibitem[{{Savonije}(2005)}]{savonije}
{Savonije}, G.~J. 2005, \aap, 443, 557

\bibitem[{{Savonije}(2013)}]{savonije2}
{Savonije}, G.~J. 2013, \aap, 559, A25

\bibitem[{{Scuflaire} {et~al.}(2008){Scuflaire}, {Th{\'e}ado}, {Montalb{\'a}n},
  {Miglio}, {Bourge}, {Godart}, {Thoul}, \& {Noels}}]{cles}
{Scuflaire}, R., {Th{\'e}ado}, S., {Montalb{\'a}n}, J., {et~al.} 2008, \apss,
  316, 83

\bibitem[{{Struve}(1955)}]{struve}
{Struve}, O. 1955, \skytel, 14, 461

\bibitem[{{Struve} {et~al.}(1957){Struve}, {Sahade}, {Lynds}, \&
  {Huang}}]{struve2}
{Struve}, O., {Sahade}, J., {Lynds}, C.~R., \& {Huang}, S.~S. 1957, \apj, 125,
  115

\bibitem[{{Townsend}(2003)}]{townsendvis}
{Townsend}, R.~H.~D. 2003, \mnras, 343, 125

\bibitem[{{Townsend}(2005{\natexlab{a}})}]{townsend}
{Townsend}, R.~H.~D. 2005{\natexlab{a}}, \mnras, 360, 465

\bibitem[{{Townsend}(2005{\natexlab{b}})}]{townsendretro}
{Townsend}, R.~H.~D. 2005{\natexlab{b}}, \mnras, 364, 573

\bibitem[{{von Zeipel}(1924)}]{vonzeipel}
{von Zeipel}, H. 1924, \mnras, 84, 665

\end{thebibliography}

\end{document}